\documentclass[10pt,aps,singlecolumn, superscriptaddress, notitlepage]{revtex4-1}
\usepackage{amsbsy}

\usepackage{amsbsy}
\usepackage{amsmath,amssymb}
\usepackage[usenames]{color}
\usepackage[colorlinks,linkcolor=black,citecolor=black]{hyperref}
\usepackage{graphicx}
\usepackage{xspace}
\usepackage{enumitem}
\usepackage{xcolor}
\usepackage{bm}
\usepackage{soul}
\usepackage{subfigure}
\usepackage{times}
\usepackage{physics}
\usepackage{qcircuit}
\usepackage{supertabular}
\usepackage{tabularx}
\usepackage{soul}
\usepackage{multirow}
\usepackage{comment}
\usepackage{textpos}
\usepackage[skins]{tcolorbox}

\usepackage{tikz}
\usetikzlibrary{positioning}
\usepackage[skins]{tcolorbox}
\newcommand{\tikzbullet}[2]{%
  \begin{tikzpicture}[baseline=-0.5ex]
    \filldraw[draw=#1, fill=#2] (0,0) circle [radius=.25em];
  \end{tikzpicture}%
}
\newcommand{\cqt}{Centre for Quantum Technologies, National University of Singapore, Singapore 117543}
\newcommand{\NQFF}{NQFF, Institute of Material Research and Engineering, A*star, 2 Fusionopolis Way, Singapore 138634}

\begin{document}

\title{A scalable narrow linewidth high power laser for
barium ion optical qubit}
\author{Morteza Ahmadi}
\email{seyedmortezaahmadi@gmail.com}
\address{\cqt}

\author{Tarun Dutta}
\email{cqttaru@nus.edu.sg}
\address{\cqt}

\author{Manas Mukherjee}
\email{manas.mukh@gmail.com}
\address{\cqt}
\address{\NQFF}
%
\begin{abstract}
The linewidth of a laser plays a pivotal role in ensuring the high fidelity of ion trap quantum processors and optical clocks. As quantum computing endeavors scale up in qubit number, the demand for higher laser power with ultra-narrow linewidth becomes imperative, and leveraging fiber amplifiers emerges as a promising approach to meet these requirements. This study explores the effectiveness of Thulium-doped fiber amplifiers (TDFAs)  as a viable solution for addressing optical qubit transitions in trapped barium ion qubits. We demonstrate that by performing high-fidelity gates on the qubit while introducing minimal intensity noise, TDFAs do not significantly broaden the linewidth of the seed lasers. We employed a Voigt fitting scheme in conjunction with a delayed self-heterodyne method to accurately measure the linewidth independently, corroborating our findings through quadrupole spectroscopy with trapped barium ions. Our results show linewidth values of  $160\pm15$~Hz and $156\pm16$~Hz, respectively, using these two methods, underscoring the reliability of our measurement techniques. The slight variation between the two methods can be attributed to factors such as amplified spontaneous emission in the TDFA or the influence of $1/f$ noise within the heterodyne setup delay line. These contribute to advancing our understanding of laser linewidth control in the context of ion trap quantum computing as well as stretching the availability of narrow linewidth, high-power tunable lasers beyond the C-band.
\end{abstract}
\maketitle
%
\section{Introduction}

In recent years, quantum computing has emerged as a promising frontier in computational technology, poised to revolutionize a wide range of industries, from finance to drug discovery \cite{herman2023quantum, blunt2022perspective, PhysRevA.106.012411}. Among the various quantum computing platforms, ion trap quantum computers have garnered significant attention due to their exceptional qubit coherence times and high-fidelity gate operations \cite{Wang2021, rudolph2022generation, Dutta2020}. However, the success of these quantum systems is profoundly sensitive to environmental factors, notably laser-induced noise, which reduces the fidelity of quantum operations. The importance of reducing laser power noise and achieving a narrow spectral linewidth in ion trap quantum computers cannot be overstated. In particular, the spectral gap between $1600$~nm to $1800$~nm is not covered by laser diodes with high power, narrow linewidth and wide tunability. Furthermore, the precise measurement of narrow linewidth has long been a fundamental challenge in various scientific and technological domains \cite{corato2023widely,lin2022electro}. High-quality resonators, such as lasers and higher-order atomic transitions, play a crucial role in numerous applications ranging from high-resolution spectroscopy to optical communications and optical clocks \cite{c:2018, wang2023linearly, piet:2015, DUTTA2019109, Dutta:16}. Accurate characterization of these sources is vital for advancing our understanding of fundamental physical phenomena of noise and enabling cutting-edge technological advancements. Traditional measurement techniques, including the use of dispersive elements like Fabry-Perot interferometers or monochromators, often encounter limitations in terms of accuracy, resolution, and measurement speed. As the demand for narrower linewidth sources continues to grow, there is a pressing need for novel measurement methodologies that can overcome these challenges and provide more precise and reliable results. \par
For narrow linewidth measurements, the most common method is the delayed self-heterodyne interferometry (DSHI), proposed by Okoshi et al. \cite{okoshi1980novel}, which enables facile measurement of linewidths in the tens of kilohertz range. This method utilizes a long fiber (several kilometers) to introduce decoherence between two laser paths. The Lorentzian linewidth can be deduced by analyzing the beat note resulting from their interference. However, a significant drawback of this method is the requirement for an excessively long fiber length, reaching hundreds of kilometers for sub-kilohertz linewidth lasers. This long fiber length is not only impractical and lossy but also introduces $1/f$ noise to the spectrum, leading to linewidth broadening and inaccurate measurements. It can also amplify nonlinear effects such as fiber-stimulated Brillouin scattering, adding an additional layer of complexity to the measurement process \cite{zhao2022narrow}. In the DSHI technique, the inclusion of short lengths of delay fiber prevents $1/f$ noise. Nevertheless, short delay fibers do not lead to a Lorentzian spectral lineshape in the beat signal. Instead, the beat signal exhibits distinct coherence envelopes. To mitigate the impact of $1/f$ noise on measurement accuracy, innovative approaches have been proposed to determine the linewidth by leveraging the coherent peaks observed in DSHI spectra (See Supplementary Material). He et al. and Xue et al. employed a coherent demodulation technique to restore the original Lorentzian line shape, enabling direct measurement of the linewidth \cite{he2019high, xue2021laser}. They achieved linewidth measurements of $2.5$~kHz and $150$~Hz, respectively. However, this approach is susceptible to demodulation errors, as it requires an initial estimation of the laser linewidth and involves iterative calculations. Another method involves utilizing specific points within the coherence envelope spectrum to establish a system of equations, with the linewidth as a variable. Huang et al. and Wang et al. adopted this approach in their respective methods: the ``contrast difference with the second peak and the second trough" and the ``dual-parameter acquisition" calculation methods. Remarkably, they achieved high-accuracy determination of laser linewidths, reaching as low as $100$~Hz \cite{huang2016laser, wang2020ultra}. These methods rely on carefully selected data points that meet specific criteria. Consequently, these are challenging to implement and prone to significant errors when applied to results with high noise levels or poorly defined data points such as servo-bumps (which are common on cavity-locked lasers) that make them inaccurate (See Supplementary Material: Fig. S2). Chen et al. proposed a Voigt fitting approach \cite{chen2015ultra}, which has demonstrated superior precision compared to directly determining the $-3$~dB linewidth value from a spectrum. This fitting scheme can partially compensate for the $1/f$ noise, making Voigt fitting a sound approach. One aspect that has hardly been touched on in all the above methods is cross-checking of the laser linewidth with other independent methods. \par
In this paper, we employ the method proposed by Chen~{\it et al.} to measure the linewidth of an external-cavity diode laser (ECDL) at $1762$~nm wavelength after being amplified by a TDFA with the ultimate goal of being used for an optical qubit gate implementation in an ion trap quantum processor. The ECDL-based master laser has a measured linewidth of about few hundred kHz. The linewidth is further reduced by locking to an ultra-stable optical cavity (USC) to an estimated linewidth of about $100$~Hz~\cite{Yum:17}. Here, we have successfully measured the linewidth of the laser after amplification and compared it to the master laser linewidth. Additionally, we validate the measured linewidth employing the atomic transition, as illustrated in ref.~\cite{Dutta2022}, in a barium ion trapped and laser-cooled using a linear Paul trap. Since diodes with high laser power and narrow bandwidth are unavailable in the spectral band $1700-1800$~nm, our study paves the way for both quantum technology and telecom technology to extend the domain of application beyond the C-band.\par
Fiber amplifiers, particularly TDFAs, play a pivotal role in the advancement of optical communications and other applications in the Near-Infrared (NIR) range. The unique properties of thulium, such as its broad emission bandwidth and compatibility with the NIR range, make TDFAs highly desirable for applications that require high-power amplification and transmission of NIR signals~\cite{walasik20231760}. By leveraging the inherent advantages of TDFAs, researchers and engineers can achieve enhanced signal quality, improved transmission distances, and higher data rates, thereby revolutionizing fields like telecommunications, fiber optic sensing, laser spectroscopy, and medical imaging~\cite{jivrivckova2023temperature, Li:23}. In this case the TDFA is crucial in extending the telecom wavelength beyond the U-band ($1625-1675$) while preserving the tuneability of a laser diode and maintaining the narrow linewidth USC-locked ECDL .   \par
This paper is structured as follows: first, we present a comprehensive overview of the working principle of TDFAs. Next, we delve into the theoretical framework and simulations of a linewidth measurement technique, which relies on fitting a Voigt profile on the beat note from the DSHI setup. Subsequently, we report the experimental measurements to determine the linewidth of the TDFA output. In order to cross-validate our measurement results, we use the output beam to implement quantum gates on a single $\rm Ba^+$ ion trapped using the  $S_\frac{1}{2}$ to $D_\frac{5}{2}$ optical transition of ion.  Finally, we discuss the viability of the linewidth measurement method and the suitability of TDFAs as laser amplifiers for ion trap-based optical quantum computing applications. 

\section{Thulium-doped fiber amplifier (TDFA) }

\begin{figure}[t]
\centering\includegraphics[width=10 cm]{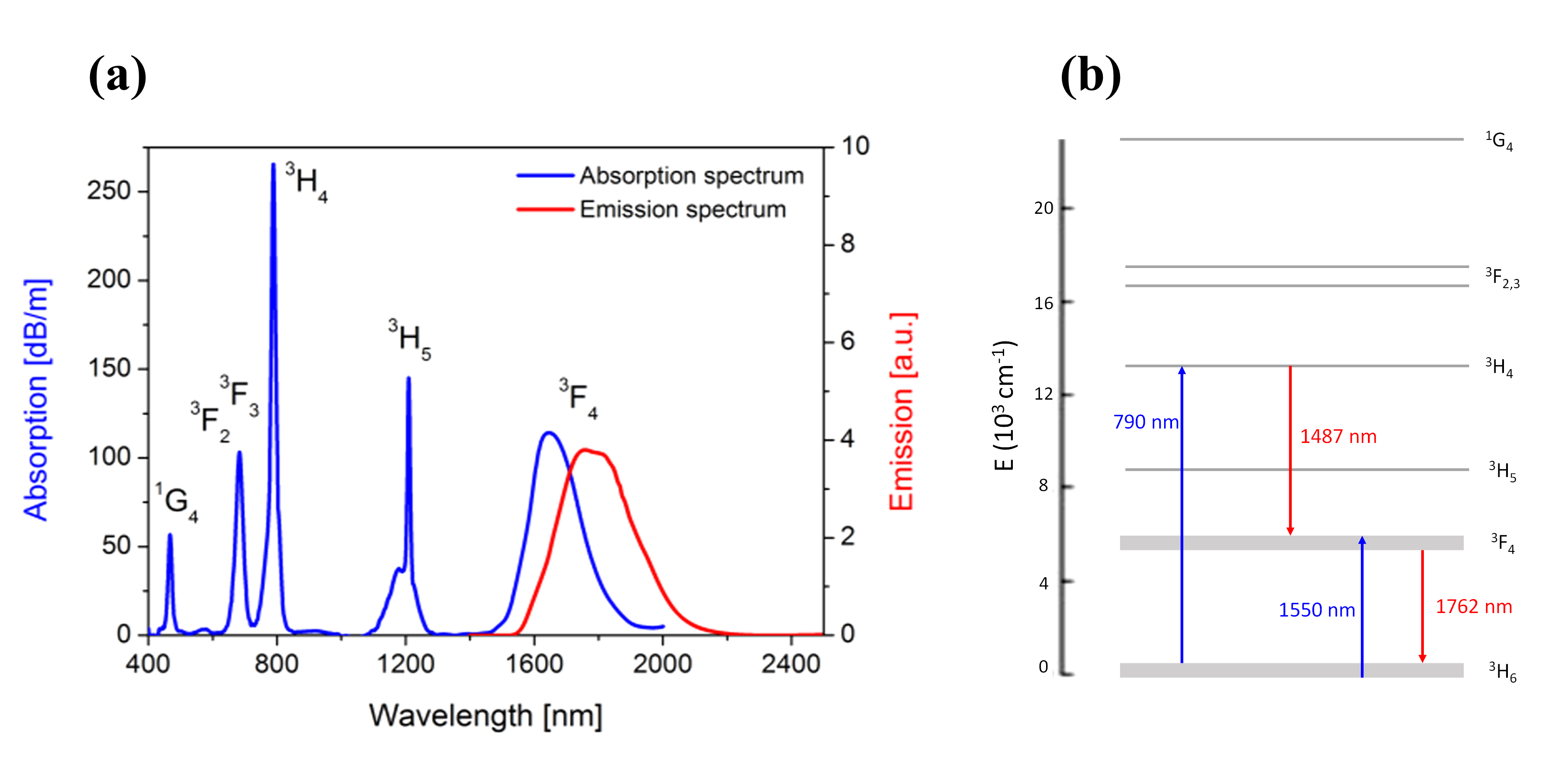}
\caption{\textbf{(a)} optical absorption and emission spectra of Tm-doped optical fiber with the indication of the corresponding absorption and emission energy levels \cite{cajzl2018thulium}, and  \textbf{(b)} its Energy-level diagram showing the two main approaches of optical pumping; in the classic method, a $790$~nm laser is used to pump the electrons to $^3\rm H_4$ state which is later decayed to $^3 \rm F_4$ by relaxation and can be used for $1762$~nm amplification. In the in-band pumping, we use a $1550$~nm laser to pump the electrons to $^3 \rm F_4$ level which can directly decay to $^3 \rm H_6$ and amplify $1762$~nm.}
\label{fig:fig1}
\end{figure}

TDFA is a fiber amplifier designed for amplifying lasers with wavelengths ranging from $1.6$ to $2.0$~$\mu$m. It achieves power amplification by introducing thulium dopant in silica fiber as the gain medium. Fig. \ref{fig:fig1}(a) depicts the continuous absorption spectrum within the $1.6-2.0$~$\mu$m interval, revealing the complex energy level structure of thulium. Specifically, this spectrum corresponds to the fine Stark manifold between $^3\rm H_6$ and $^3\rm F_4$, which is indicated by two bold lines. By optical pumping the silica fiber doped with $\rm Tm^{3+}$ ions, the electrons transit to metastable states. Once a sufficient power level is reached to achieve population inversion, the excited ions amplify the seed light through stimulated emission~\cite{balda2007spectroscopy}.
To amplify the $1762$~nm laser, two main approaches for pumping atoms to excited states are employed. Fig. \ref{fig:fig1}(b) illustrates the traditional three-energy off-band pumping and the in-band pumping methods. In the case of off-band pumping, a $790$~nm laser diode is used to pump electrons to the $^3\rm H_4$ state, followed by their relaxation to the metastable energy levels of  $^3 \rm F_4$. The transition from  $^3\rm F_4$ to $^3 \rm H_6$ is selected for stimulated radiation, leading to power amplification. The unlabelled energy levels in the vicinity of $^3 \rm F_4$ in Fig. \ref{fig:fig1}(b) provide a broader gain range around $1800$~nm. However, it is also possible to amplify the laser using only the two Stark manifolds, $^3\rm F_4$ and $^3\rm H_6$. By pumping $\rm {Tm}^{3+}$ ions with a $1550$~nm laser to the high energy manifolds of $^3\rm F_4$, rapid thermalization occurs within each manifold within picoseconds. Consequently, at low temperatures such as room temperature, the majority of ions reside in the lower sub-levels, making them inaccessible to pump light. The seed light, generated by stimulated emission, then propagates back to $^3\rm H_6$ without intermediate manifolds, thus amplifying the $1762$~nm seed laser. In our experiment, the TDFA is in-band pumped. The main contributor to the linewidth broadening of the amplified laser is spontaneous emission. However, if the seed light intensity is sufficiently high to extract the electrons in the upper energy level, the power of amplified spontaneous emission is suppressed, leading to a narrower output laser linewidth.
\section{Voigt Fitting Method}
It is shown by Scully and Lamb~\cite{scully1967quantum} that the spectral distribution of the lasers follows a Lorentzian profile due to the white noise present in their spectrum. In the DSHI method, a Gaussian profile is also added to the spectral distribution due to the addition of the delay fiber, which introduces flicker ($1/f$) noise. It is shown by Mercer \cite{mercer19911} that the DSHI spectrum of the lasers demonstrates a Voigt profile, which is the convolution of the Lorentzian and Gaussian lineshapes  and is expressed as:
\begin{equation}
V(f)=\int_{-\infty}^{+\infty} G(f')L(f-f') \, df'
\end{equation}
in which $G(f)$ and $L(f)$ are respectively the normalized Gaussian and Lorentzian lineshapes defined as:
\begin{equation}
G(f)={\frac{2\sqrt{\ln{2}}}{\sqrt{\pi}\Delta f_G}}\exp[\frac{-4\ln2(f-f_0)^2}{\Delta f_G^2}]
\end{equation}
\begin{equation}
L(f)={\frac{\Delta f_L}{2\pi}}\frac{1}{(f-f_0)^2+\frac{\Delta f_L^2}{4}.}
\end{equation}
In equations (2) and (3), $\Delta f_G$ and $\Delta f_L$ represent the Gaussian and Lorentzian linewidths, respectively, and $f_0$ is the frequency shift between the two paths in the DSHI method, which is generated by either an acousto-optic modulator (AOM) or electro-optic modulator (EOM).  The $\Delta f_G$ and $\Delta f_L$ can be estimated by fitting a Voigt profile on the  beat note acquired from the DSHI method. Numerous  algorithms have been proposed to precisely estimate the Voigt parameters, accommodating various combinations of Lorentzian and Gaussian contributions. These fitting techniques involve the utilization of the Voigt function and its derivatives for every data point in the spectrum, iterating until a minimal error is achieved \cite{zheng2022comparison}.
In our measurements, we have chosen to employ the method proposed by Chen et al. that demands fewer computational resources when compared to nonlinear least squares fitting procedures. This approach approximates the relation between the Voigt spectrum and the Lorentzian and Gaussian spectra through the following expression:
\begin{equation}
\Delta f_V=\frac{1}{2}(1.0692 \Delta f_L+\sqrt{0.866639\Delta f_L^2+4\Delta f_G^2})
\end{equation}
in which $\Delta f_V$ is the Voigt linewidth. As the broadening  of the $1/f$ noise tends to be mostly noticeable near the central frequency, the $3$~dB width of the spectrum is significantly influenced by the Gaussian component. Conversely, the $20$~dB width is primarily governed by the Lorentzian contribution. This value serves as the initial estimate for the Lorentzian linewidth. Subsequently, we estimate the Gaussian component using the $3$~dB linewidth and Eq. (4). With these estimated values for the Gaussian and Lorentzian linewidths we can construct a Voigt profile using Eqs. (1)-(3). By comparing the Voigt profile to the measured spectrum, we iteratively refine our estimate of the Lorentzian linewidth, seeking the value that yields a 20-dB width for the Voigt profile matching the measured width. Notice that in the DSHI method, the measured Lorentzian linewidth is the combined linewidth of two lasers with similar Lorenztian linewidths and the actual linewidth of the laser under test will be half of the measured value if the linewidths are the same~\cite{canagasabey2011comparison, bai2021narrow}.

\section{Experimental Analysis}

\begin{figure}[t]
\centering\includegraphics[width=0.9\textwidth]{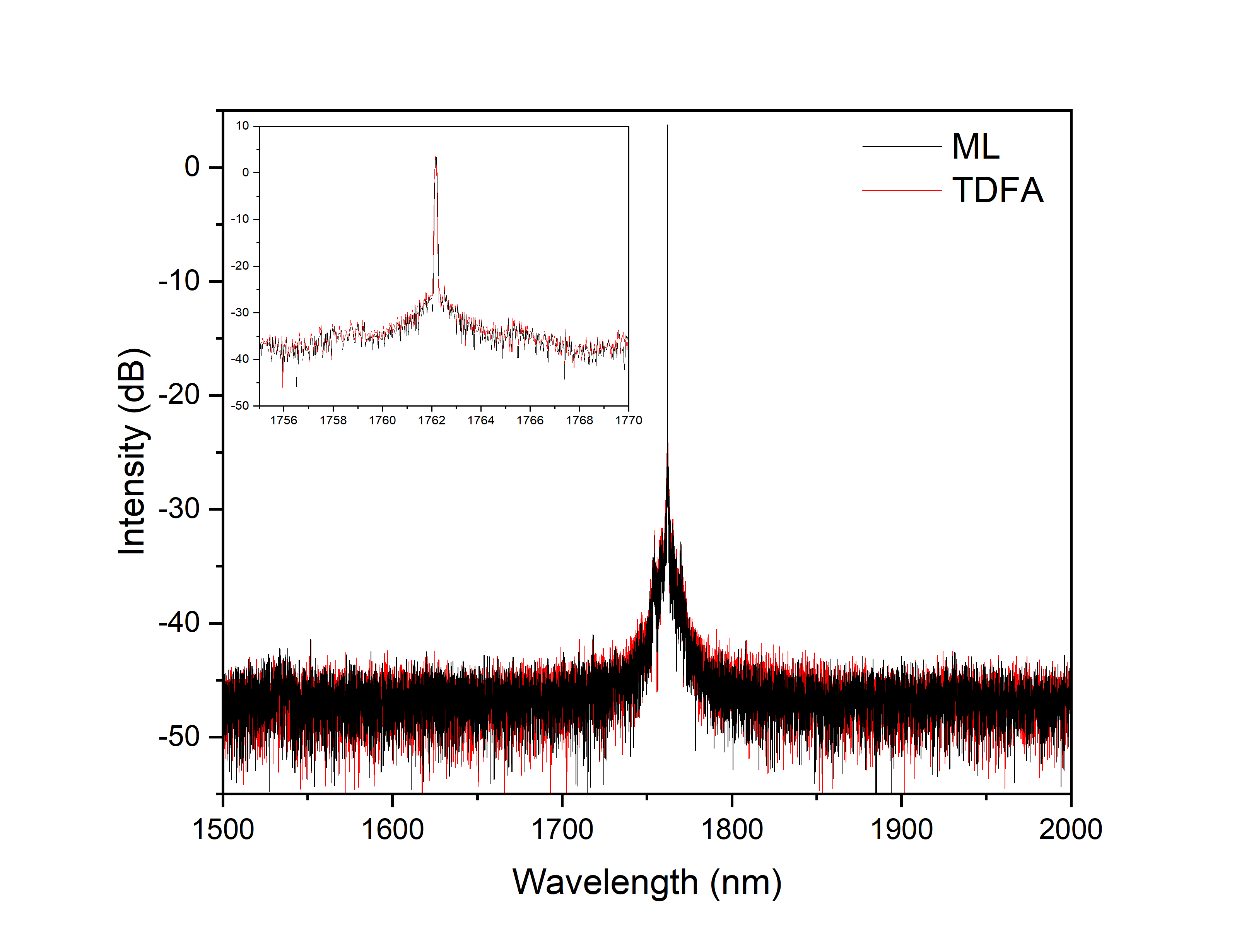}
\caption{
The spectra of the master laser and TDFA were analyzed using an optical spectrum analyzer (OSA), with a zoomed-in view (inset) around $1762$~nm. The OSA boasts a resolution of $\pm0.0002$~nm at $1000 $nm and an optical rejection ratio exceeding $40$~dB.
}
\label{fig:fig2}
\end{figure}

\subsection{Optical Spectrum Analyzer (OSA)}

We employed a $771$~series laser spectrum analyzer from Bristol instruments to characterize the master laser and TDFA spectra, which are illustrated in Fig.~\ref{fig:fig2}. Both the master and fiber lasers demonstrate an almost identical spectral behavior, with the TDFA showing a slightly higher noise floor which can be attributed to the amplified spontaneous emission (ASE). Both lasers demonstrate an optical signal-to-noise ratio of more than 50 dB in the current measurement which the actual value might be even higher due to the limited intensity resolution of our OSA. Furthermore,  since the instrument's spectral resolution is around 2 GHz, we can not estimate the expected sub-kilohertz linewidth of our lasers using this direct spectral measurement method.

\subsection{DSHI Method}

\begin{figure}[t]
\centering
\includegraphics[width=0.9\textwidth]{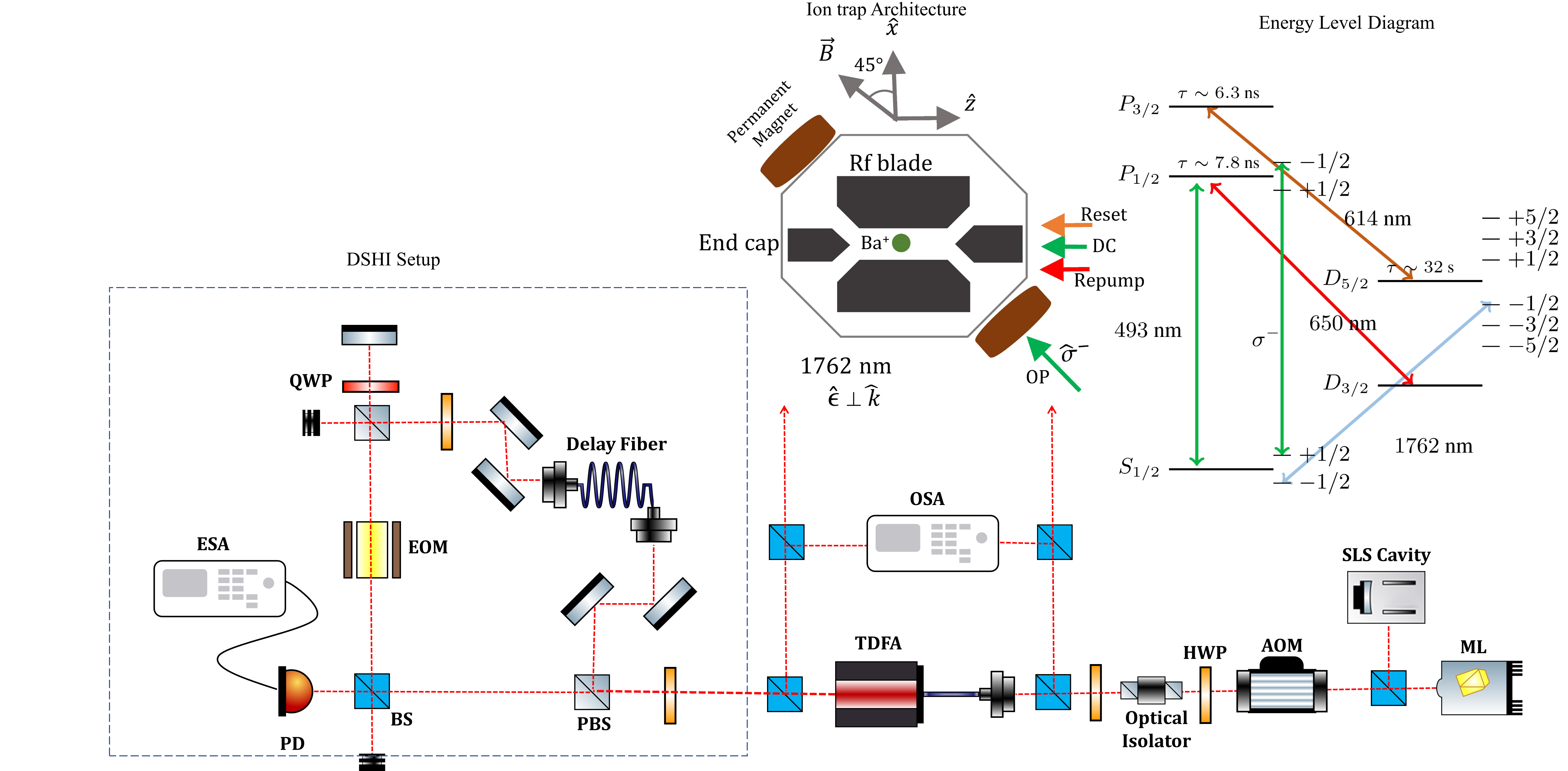}
\caption{The schematics of the optical setup used for linewidth measurement: $10$~mW of light from a frequency-stable master laser (ML) is injected into the Thulium-doped fiber amplifier (TDFA) through an optical isolator, utilized to minimize optical feedback. The frequency of the master laser is phase-locked to a high-finesse stable cavity. Both lasers are actively monitored using a Bristol wavemeter, and their spectral quality is compared using a Bristol optical spectrum analyzer (OSA). A small fraction of the fiber laser's output power is dedicated to linewidth measurement using a DSHI setup (dashed line), while the remaining power is directed to the ion trap setup. The portion of light from the fiber amplifier designated for the heterodyne setup is split into two paths using a polarizing beam splitter (PBS) and a half-waveplate (HWP). One path passes through a delay fiber and an electro-optic modulator (EOM), while the other part goes directly through a beam splitter (BS) to reach the photodiode. The photodiode (PD) signal is observed using an electronic spectrum analyzer (ESA). AOM: acousto-optic modulator, QWP: quarter waveplate, SLS cavity: stable laser system cavity. 
In the top right of the figure, the energy level diagram of a barium ion is depicted: $493$~nm (DC) and 650 nm (Repump) lasers are utilized for Doppler cooling, while $1762$~nm is employed as a quadrupolar qubit laser. The $614$~nm laser, known as the reset laser, is used to de-shelve the population from the D$_{5/2}$ state. The top left of the figure illustrates our ion-trap architecture: RF blades confine the barium ion radially, while end cap electrodes confine the ion along the axial direction, termed as the trap axis (z-axis). The schematic also indicates the direction of all lasers directed toward the trap, including the direction of the magnetic field. A uniform magnetic field at the trap center is generated by a pair of permanent magnets of a low-temperature coefficient. The Doppler cooling laser is aligned along the trap axis, while the qubit laser is injected perpendicular to the trap axis but makes an angle of 45 degrees with the quantization axis. See the text for details.
}
\label{fig:fig3}
\end{figure}

\begin{figure}[b!]
\centering\includegraphics[width=12 cm]{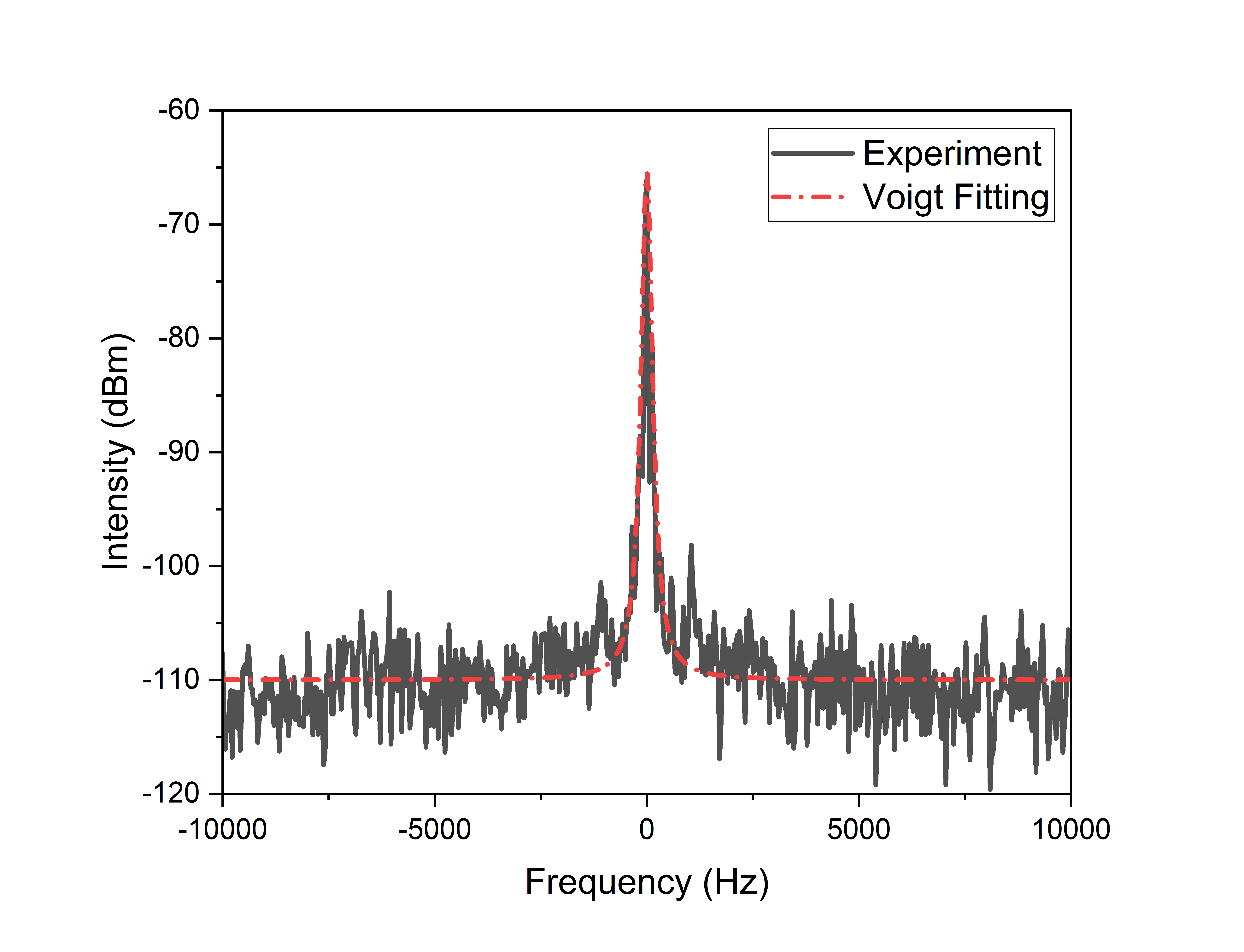}
\caption{The beat note acquired from the DSHI method and the fitted Voigt profile at 170126.4 THz frequency show a linewidth of $160\pm15$ Hz. For the beat notes acquired at other frequencies see Supplementary Material S5.}
\label{fig:fig4}
\end{figure}

Fig.~\ref{fig:fig3} schematically demonstrates the DSHI setup used in our study. An ultra-stable ECDL is used to trigger the stimulated emission in the TDFA. The amplification power is controlled by the pump current and can achieve optical power outputs of over $500$~mW (Figure S.3). The measurements in this experiment are conducted by employing ~$10$ mW optical power of the master laser and 1 A electrical current input, which yields $\sim$$150$~mW TDFA output. The TDFA output laser beam is split into two paths. In the first path, the laser travels directly to the photodetector following polarization adjustment while in the other path, the laser is delayed by a $5$~km long single-mode fiber. In this path, a $7$~MHz resonant electro-optic phase modulator is used to create frequency sidebands. The polarization of the laser in the second path is adjusted to match with that of the first path. The delayed line is used to partially destroy the coherence between the two arms, and its length is determined by the estimated coherence time of the laser under test. Subsequently, the interference pattern is generated on the photodetector and can be observed using a high-resolution electronic spectrum analyzer (ESA). Since the ESA provides a better resolution down to $1$~Hz, this heterodyning technique allows down-conversion of the beat frequency with high resolution and is better suited than OSAs to measure the narrow linewidths. The heterodyne spectrum of the measurement along with the fitted Voigt profile using the fitting method explained in section 3,  is shown in Fig. \ref{fig:fig4}.  The measured combined Lorentzian linewidth is estimated to be around $320$~$\pm$ $30$ Hz, giving a single laser linewidth of $160$~$\pm$ $15$ Hz. This assumes that the laser linewidth in the two arms are the same, which in this case is a valid assumption. This measured linewidth warrants a cross-check with other alternative methods. Since the final application of the laser is to drive the optical qubit, in the following, we should use an alternative means to cross-check the linewidth via high-resolution optical spectroscopy as well as gate fidelity analysis.
 
\subsection{ High resolution atomic spectroscopy}
Here we have used a single trapped and laser-cooled barium ion to perform an optical quadrupole spectroscopy. Similar to heterodyne measurement, quadrupole spectroscopy is an advanced technique used to measure the linewidth and spectral properties of ultra-narrow linewidth lasers with high precision. In our trapped ion setup as illustrated in Fig. \ref{fig:fig3}, (for a more detailed illustration, see ref.~\cite{Dutta2022, Yum:17, Dutta2020, PhysRevA.106.012411}), single barium ion $^{138}$Ba$^+$ is confined in a linear RF Paul trap. The trapping potential is well approximated as harmonic with a secular axial frequency of $ \omega_z = 0.5$ MHz and two nearly degenerate radial frequencies of $\omega_r = 1.5$ MHz.  A uniform homogeneous magnetic field splits the ground states by a frequency of $\sim 5$~MHz. Our optical qubit is encoded in the ground state  $6S_{\frac{1}{2}}$ and the metastable state 5$D_{\frac{5}{2}}$, where the natural lifetime of the D$_{\frac{5}{2}}$ state ($\sim 30$~s ) sets the de-coherence time (T1) of the optical qubit. Since the natural linewidth of the D-state is mHz, it is a perfect reference to measure laser linewidth down to Hz. In this experiment, the direction of $1762$~nm laser light propagation relative to the quantization axis defined by the magnetic field direction limits the allowed strong transitions between  S$_{\frac{1}{2}}$ and D$_{\frac{5}{2}}$ manifolds to $\Delta m = \pm 0, 2$ only. We have used $\Delta m = 0$ as our qubit because it is the least sensitive to magnetic field fluctuations. It is essential to acknowledge that fluctuations in the magnetic field can introduce linewidth broadening, which can impact the performance of our quantum system as discussed in ref.~\cite{Dutta2022}. To mitigate these effects, we have incorporated a low-temperature coefficient permanent magnet into our setup. This type of magnet helps stabilize the magnetic field, ensuring the reliability and precision of our spectroscopic measurements and quantum operations.
\begin{figure}[t!]
     \begin{center}
        \subfigure[ ]{%
            \label{fig:fig5a}
            \includegraphics[width=0.5\textwidth]{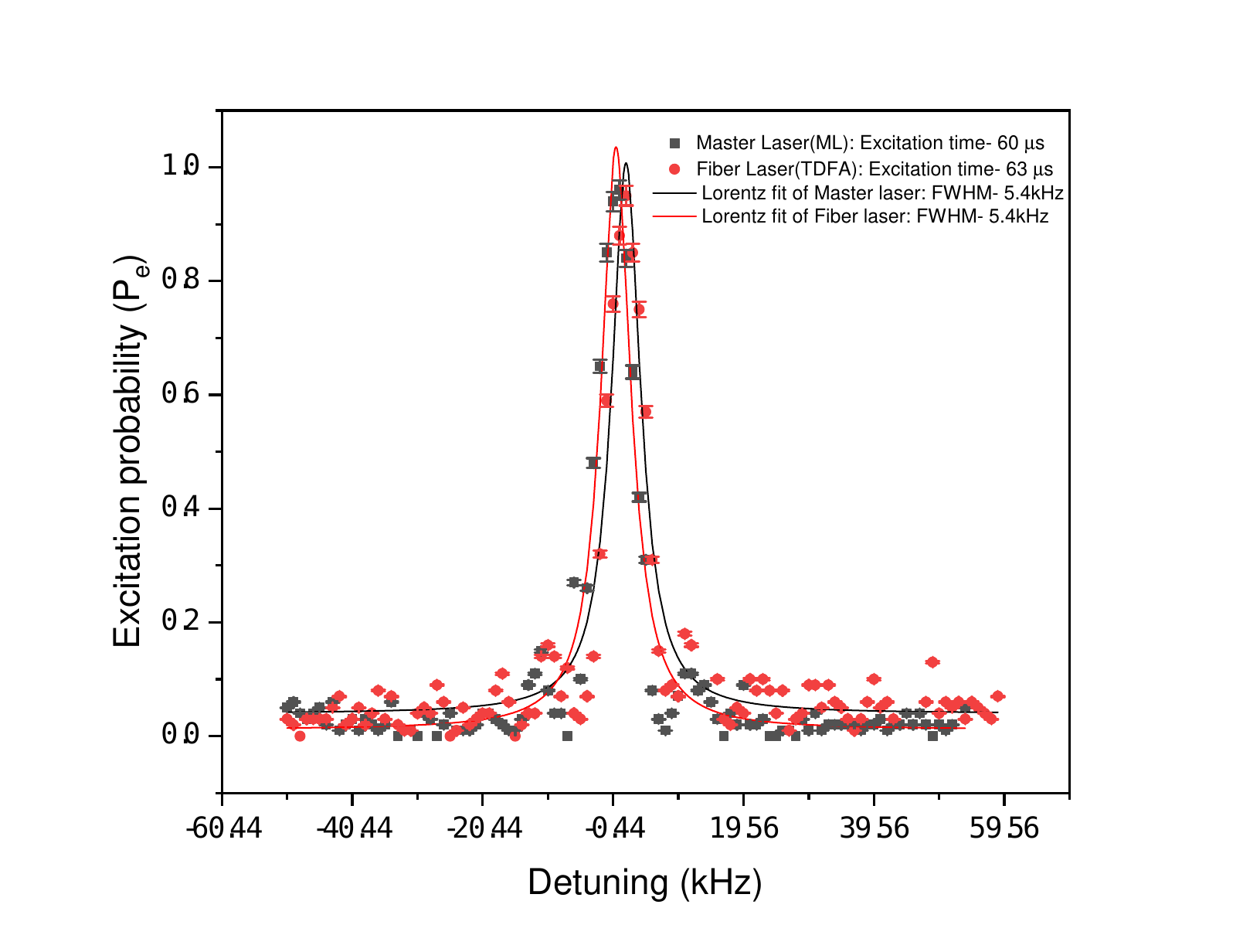}
        }%
        \subfigure[ ]{%
           \label{fig:fig5b}
           \includegraphics[width=0.5\textwidth]{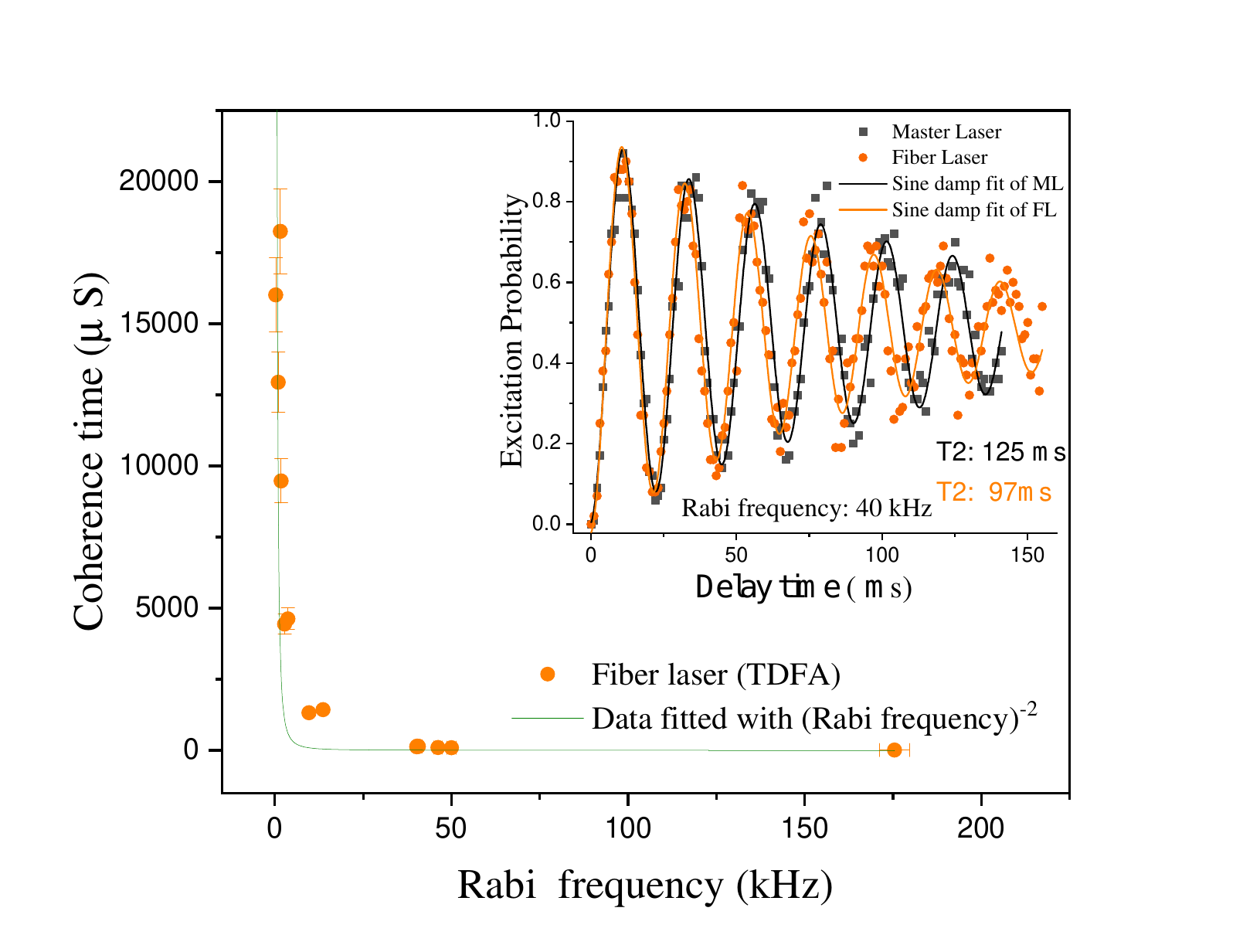}
        } 
       
    \end{center}
\caption{The resonance spectra and the Rabi oscillations between ML and TDFA: (a) A spectrum near the carrier transition of the $6S_{\frac{1}{2}}(m_j=-\frac{1}{2})$ to  $5D_{\frac{5}{2}}(m_j=-\frac{1}{2})$ of a single barium ion for both the lasers is shown in the same scale for comparison. When fitting the data with Lorentzian profiles, it is evident that both spectra exhibit a linewidth of $5.4~\pm 0.3$~kHz. Notably, the length of the probe pulse for the master laser appears marginally shorter when compared to the fiber laser. (b) The coherence time is plotted against the Rabi frequency (an alternative representation of the intensity) of a single ion using the fiber laser. The data is described by a fit proportional to the inverse square of the Rabi frequency~\cite{PhysRevA.57.3748}. As the fiber laser's power diminishes, there's a notable rise in coherence time. The inset graph contrasts the Rabi Oscillations obtained by using both the master ($\blacksquare$) and fiber lasers ($\protect\tikzbullet{orange}{orange}$) with the duration of the probe pulse kept constant at about $ 12\mu$s (Rabi frequency $\sim 40 $ kHz). Fitting the data using a damped sine function reveals details about the coherence time. The plot indicates that the coherence time is slightly shorter for the fiber laser as compared to the master laser indicating higher intensity noise.
}
   \label{fig:fig5}
\end{figure}
\begin{figure}[t]
\centering
\includegraphics[width=0.9\textwidth]{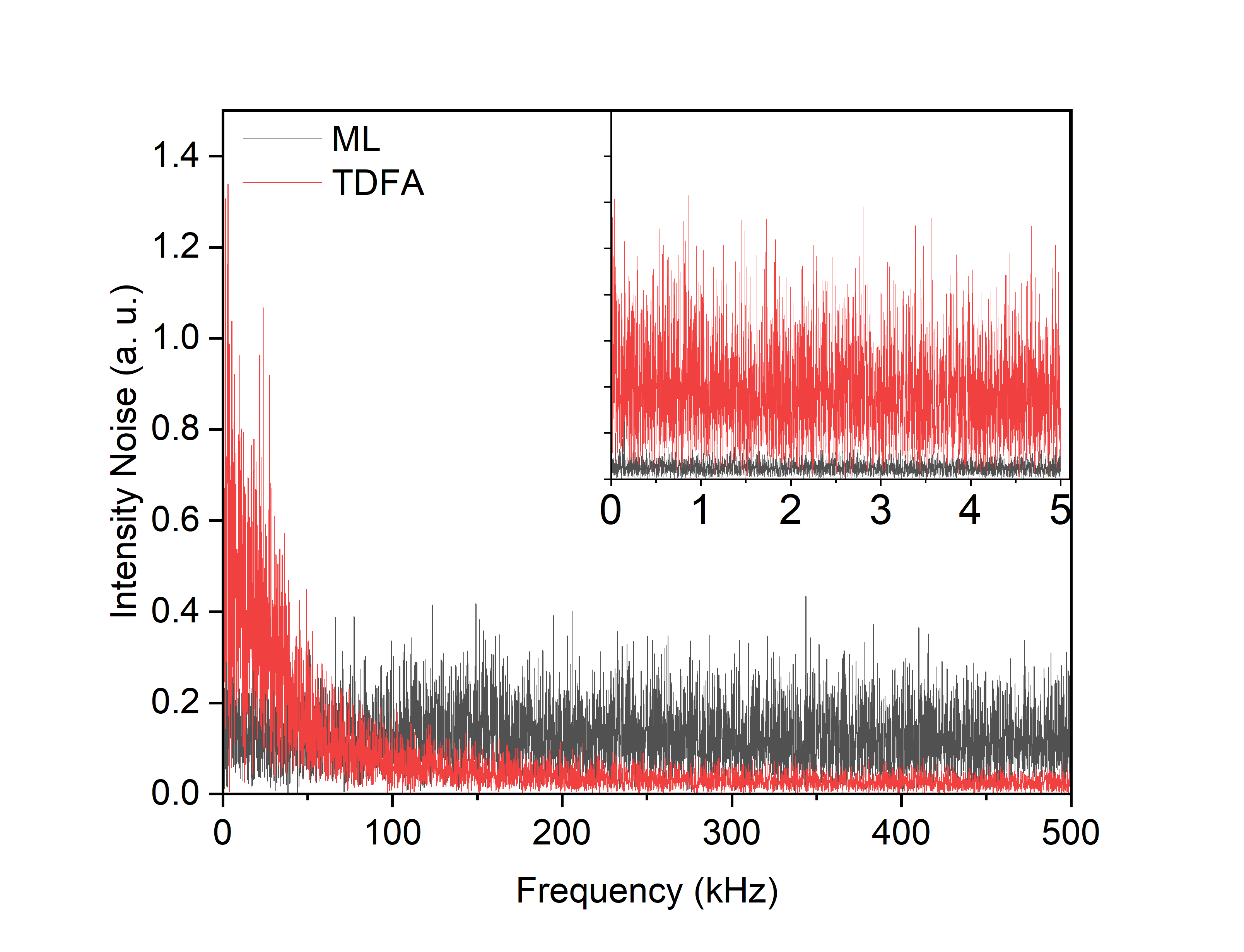}

\caption{Intensity fluctuations of the master laser and TDFA in frequency domain measured using a wideband photodetector showing a higher noise floor for the TDFA laser, particularly in the lower frequency range. Since the two lasers demonstrate the same frequency linewidths, the slight difference between the coherent behavior of the two lasers can be attributed to their intensity noise. (for the long-term intensity behavior of the laser one can see the Supplementary Material Fig. S6) }
\label{fig:fig6}

\end{figure}

The linewidths of both the master and TDFA lasers are measured by performing high-resolution spectroscopy on the quadrupole transition $6S_{\frac{1}{2}}(m_j=-\frac{1}{2})$ to  $5D_{\frac{5}{2}}(m_j=-\frac{1}{2})$ with a single trapped and laser-cooled barium ion. The master laser is stabilized to an ultra-high-finesse $(10^{6})$ reference cavity. A comprehensive description of our narrow linewidth master (diode) laser system can be found in ref~\cite{Yum:17, Dutta2022}. The resulting spectra of this quadrupole transition for both lasers are presented in Fig.~\ref{fig:fig5}(a). These spectra were obtained by adjusting the duration of the probe pulse to achieve the same linewidth when fitting them with a Lorentzian function. Both lasers exhibit a linewidth of $5.4$ kHz as shown in Fig.~\ref{fig:fig5}(a) with $3\mu$s difference in the probe pulse length. This discrepancy provides valuable insights into the Rabi frequency difference between the master and fiber laser, which amounts to approximately 1 kHz.
To attain an identical linewidth, the need for a higher Rabi strength in the fiber laser compared to the master laser hints at the spectral quality discrepancy between the two lasers. This disparity suggests that the spectral quality of the fiber laser may be inferior to that of the master laser, possibly due to a slight increase in ASE in the fiber laser or the presence of other factors unique to the fiber laser system. To validate the observed difference in spectral quality, we conducted Rabi oscillation measurements with an identical probe pulse set up for both lasers.

Measurements of Rabi oscillations using both master laser ($\blacksquare$) and fiber laser  ($\tikzbullet{orange}{orange}$) on the carrier $6S_{\frac{1}{2}}(m_j=-\frac{1}{2})$ to  $5D_{\frac{5}{2}}(m_j=-\frac{1}{2})$ quadrupole transition for a thermal motional state is presented in inset of Fig.~\ref{fig:fig5}(b).  On the carrier, the ion oscillates periodically between the ground state and the excited state. The decay of the oscillation shown in the inset of Fig.~\ref{fig:fig5}(b) is due to the decoherence which degrades the fidelity of these oscillations. However, various factors can introduce decoherence. Here, the loss of contrast is primarily attributable to the de-phasing of the atomic qubit. The solid line in Fig. \ref{fig:fig5}(b) represents a fit to a sine function with decaying amplitude. This fitting procedure helps to quantify the decoherence time for both lasers. Notably, the decoherence time, as observed from the Rabi oscillation, for the fiber laser is slightly shorter than that of the master laser, even with the same probe pulse duration. This can be attributed to several factors. One of the reasons we suspect is that fiber lasers may experience more phase noise or fluctuations in the laser phase due to the properties of the optical fiber, interactions with the environment, or instabilities in the fiber laser's components. This phase noise can lead to faster de-phasing and a little shorter coherence time compared to the master laser.  Furthermore, we have observed discrepancies in spectral purity between the master laser and the fiber laser (TDFA), which have manifested as disparities in their respective coherence times as seen in the figure. Reducing the power of a fiber laser increases the coherence time, demonstrated in Fig. \ref{fig:fig5}(b). The data is described by a fit proportional to the inverse square of the Rabi frequency~\cite{PhysRevA.57.3748}. When the power of the fiber laser is reduced, it often corresponds to a decrease in the linewidth of the laser emission due to the reduction in spontaneous emission.  In a laser system, a narrower linewidth is associated with a more monochromatic and coherent output. Therefore, as the power decreases and the linewidth narrows, the coherence time tends to increase.

Another reason that can speed up the qubit decoherence is the intensity noise. In a TDFA, the amplification process generates inferior output stability than the master laser, and this can be intensified when the TDFA input is modulated using an AOM. In this situation, the trapped ion will receive a higher intensity noise than when it is directly addressed by the master laser, which might result in a shorter decoherence time. To investigate the intensity noise on the master and TDFA lasers, we used a Thorlabs photodetector to monitor their power stability with their results in the frequency domain is shown in Fig.~\ref{fig:fig6}. It can be seen from the measurement that the TDFA possesses a higher level of intensity noise especially at lower frequencies compared to the master laser which may create faster decoherence. This higher level of intensity noise in fiber lasers compared to ECDLs has been previously shown \cite{kapasi2020tunable}. Since the linewidth of the laser is determined by its phase noise and with the same $\pi$~time, both lasers are showing the same linewidths; we can, therefore attribute the shorter decoherence time of the TDFA with respect to the master laser to its higher intensity noise.   \par
\begin{figure}[ht!]
\centering\includegraphics[width=12 cm]{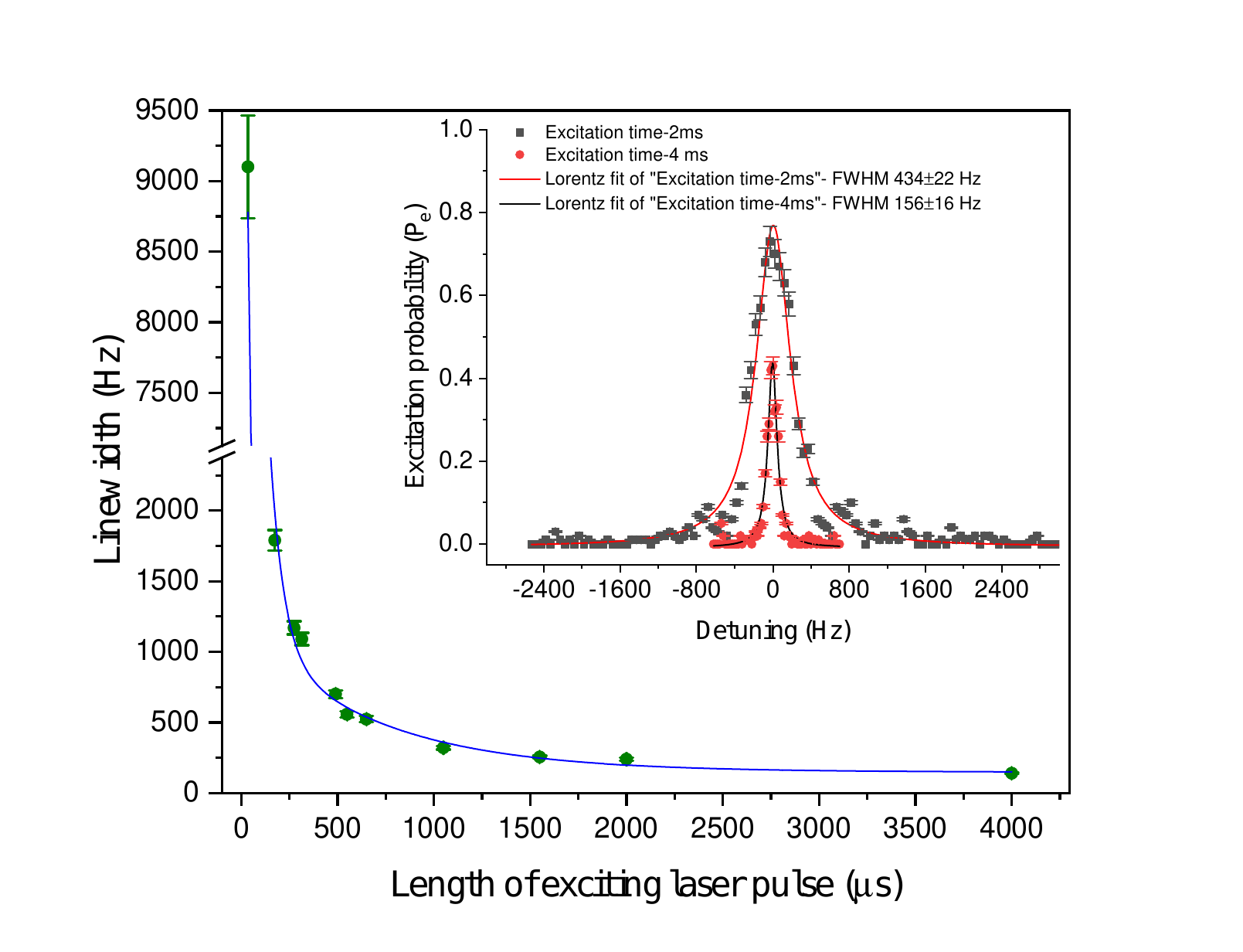}
\caption{The figure depicting linewidth versus the length of the excitation laser pulse provides valuable insights into the behavior of a fiber laser's spectral characteristics. Specifically, it reveals the upper limit of the linewidth achievable in a fiber laser under varying pulse durations.  The data is fitted (blue solid line) with an inverse relationship to the pulse duration. This implies that as the pulse duration decreases, the linewidth tends to increase, aligning with the principles of power broadening. This observation becomes particularly relevant in the realm of pulsed coherent excitation. (Inset) A spectrum of carrier transition of the $6S_{\frac{1}{2}}(m_j=-\frac{1}{2})$ to  $5D_{\frac{5}{2}}(m_j=-\frac{1}{2})$ of a single barium ion: This is recorded with a high-resolution scan of carrier transition with a long excitation time of 2 ms ($\blacksquare$) and 4 ms ($\protect\tikzbullet{red}{red}$) was used. The Lorentzian fitting of the spectrum yields a linewidth measurement of $434$$\pm 22$ Hz and $156\pm 16$ Hz for 2-ms and 4-ms excitation pulses. }
\label{fig:fig7}
\end{figure}
%

In high-resolution spectroscopy with a fiber laser, reducing the power of the laser can be an effective means to narrow the linewidth of the spectral line under study. This spectral linewidth, determined during high-resolution scans, effectively sets an upper limit on the linewidth of the exciting laser. Fig. \ref{fig:fig7} showcases the trade-off between the laser power and pulse duration in affecting the spectral linewidth in high-resolution spectroscopy. The data exhibits an inverse correlation with pulse duration. This effectively illustrates the linewidth decreases as the power of the laser is reduced, indicating an increase in the length of the excitation laser pulse, aligning with the principles of power broadening. This observation becomes particularly relevant in the realm of pulsed coherent excitation~\cite{2001OptCo.199..117V}. The information regarding the linewidth at each length of the exciting laser pulse is obtained through Lorentzian fitting of the carrier spectrum- the spectra of the quadrupole transition, as discussed before. The high-resolution scanning of the carrier spectrum, along with their respective fittings, for pulse durations of $2$ and $4$ ms, is depicted in the inset of Fig. \ref{fig:fig7}. It's worth noting that these spectra are acquired under conditions of exceptionally low optical intensity at the ion site, employing exciting laser pulses with a duration of $2 $  ($4$ ) ms, yielding a Rabi frequency of approximately $500 ($$ 250)$ Hz. The spectral lineshape is fitted using the Lorentzian function, which provides a comprehensive characterization of the linewidth for the lasers. Within this context, the linewidth precisely corresponds to the FWHM of the Lorentzian fit. Our measurements indicate a linewidth estimate for the fiber laser, with an upper limit of $156$ $\pm$ $16$ Hz, all while maintaining a Rabi frequency of 250 Hz. Notably, the peak height of transition is reduced, as shown in the inset of Fig. \ref{fig:fig7}, with the increase of excitation time because the transition was de-saturated by reducing the laser power to less than 0.5 $\mu$w. This reduction in the peak height imposes a limitation, preventing us from extending the excitation time beyond 4 ms. \par

Due to the wide absorption band of Thulium, we expect the TDFA to exhibit amplification in a broad NIR range. In order to investigate the tunability of the TDFA, we locked the ECDL to the USC with central frequencies at $170124.9$ and $170123.4$~THz which are respectively $1.5$ and $3.0$~GHz away from the resonance frequency used for the ion qubit resonance. The beat notes acquired for these cases and their fitted Voigt profiles are illustrated in Fig. S5.  As we can see from the figure, in both cases, the linewidths remain the same around $160$~Hz, which confirms the capability of the TDFA to amplify the input light in broad NIR range without altering its noise profile and linewidth. 

\section{Discussion}
We introduced the in-band pumped TDFA as a capable tool to achieve high amplification in the NIR wavelength region while maintaining the master laser linewidth and tuneability, ideal for quantum information processing. For a detailed analysis of TDFA performance with regard to power and polarization, one can refer to previous studies \cite{walasik20231760}. We employed the Voigt fitting scheme proposed by Chen et al. \cite{chen2015ultra}, using a DSHI setup to measure the linewidth of our TDFA while it is being seeded by an ultra-high finesse ECDL laser. Compared to other methods for measuring ultra-narrow linewidths, the Voigt fitting method demonstrated significant advantages. For instance, the contrast difference between the second peak and the second trough method \cite{huang2016laser} required accurate measurement of the intensities between the second peak and trough, which was not possible because they fell within the servo bandwidth of our PID controller, and the mentioned peaks were buried inside the bumps (Detailed description of the method is provided in the Supplementary Materials). Furthermore, we could not use the other coherent envelopes on our DSHI spectrum outside the servo bandwidth as they would lack sufficient accuracy for our measurements \cite{li2019laser}. Although, in theory, we could have avoided the coherent envelopes in the heterodyne spectra either by increasing the delay line length or by using another narrow linewidth laser as the local oscillator for direct measurement of the linewidth, in practice, for linewidths as narrow as 100 Hz, this requires hundreds of kilometers of fibers or an identical or better second laser source which is not readily available. Nevertheless, the Voigt fitting method accurately estimated the linewidth by considering only the central unperturbed peak of the DSHI beat note, later verified by quadrupole spectroscopy.

The quadrupole spectroscopy performed with the same laser on the mHz transition line reconfirmed the heterodyne measurement results. It also showed that the TDFA does not introduce any significant linewidth broadening over the seed laser linewidth even at the level 100 Hz linewidth. Furthermore, the quadrupole spectroscopy showed that the qubit demonstrates a higher coherence time when addressed by the ECDL directly compared to the TDFA, and the latter might slightly reduce the fidelity of the qubit in similar $\pi$-times. Considering this, along with the similar linewidths of the lasers, we can attribute the faster decoherence in the TDFA to its intensity noise, which might be due to the nonlinear amplification processes or the ASE in TDFA. As shown by Day et al.,~\cite{day2022limits} intensity noise can be a limiting factor for lasers used to manipulate qubits preventing them from reaching their spontaneous emission noise floor.  It is important to emphasize that while the TDFA exhibits a faster decoherence when compared to the master laser, this observation does not diminish its suitability for qubit manipulation as the increase is only marginal and can be removed by intensity stabilization. Therefore, the TDFA remains a viable option for qubit manipulation, offering versatility in quantum computing applications particularly solving the problem of scaling. As the number of qubits increases the requirement of high power of the narrowband laser is an essential component. Furthermore, it is shown that the TDFA exhibits a high bandwidth in NIR range which makes it an excellent tool for applications in telecom wavelengths. 

In summary, our research introduced the in-band pumped TDFA as a powerful tool for NIR wavelength amplification while preserving the linewidth, polarization and tunability of the master laser down to 100 Hz. We focused on investigating the TDFA's linewidth broadening and noise profile. Our findings indicate a slight increase in intensity noise and decoherence in the TDFA compared to the ECDL. The difference is negligible as compared to the benefits of scalability and tuneability offered by the amplifier for the purpose of quantum gate operations with high fidelity and also for applications in telecom wavelegths.

\bibliography{References}
\newpage
\pagebreak
\widetext
\begin{center}
\textbf{\large Supplementary Materials: High power narrow linewidth laser for scaling barium ion optical qubit.}
\end{center}
\setcounter{equation}{0}
\setcounter{figure}{0}
\setcounter{table}{0}
\setcounter{page}{1}
\makeatletter
\renewcommand{\theequation}{S\arabic{equation}}
\renewcommand{\thefigure}{S\arabic{figure}}
\renewcommand{\bibnumfmt}[1]{[S#1]}
\renewcommand{\citenumfont}[1]{S#1}
 

\section{Extracting the linewidth from the coherent envelopes }

The power spectral density (PSD) function of the beat note (\(S\)) can
be modeled by the following equations~\cite{richter1986linewidth, huang2016laser}:
\begin{align*}
S(f) &= S_{1}{\times S}_{2}{+ S}_{3}\\
S_{1} &= \frac{{P_{0}}^{2}}{4\pi}\frac{\mathrm{\Delta}f}{{\mathrm{\Delta}f}^{2} + {(f - f_{EOM})}^{2}}\\
S_{2} &= 1 - e^{( - 2\pi t_{d}\mathrm{\Delta}f)}\left\lbrack \cos\left( 2\pi t_{d}\left( f - f_{EOM} \right) \right) + \mathrm{\Delta}f\frac{sin(2\pi t_{d}\left( f - f_{EOM} \right))}{f - f_{1}} \right\rbrack\\
S_{3} &= \frac{\pi{P_{0}}^{2}}{2}e^{( - 2\pi t_{d}\mathrm{\Delta}f)}\delta\left( f - f_{EOM} \right)
\end{align*}

in which \(f\) is the measurement frequency, \(P_{0}\) stands for the
total optical power received by the photodetector, \(\mathrm{\Delta}f\)
shows the linewidth of the TDFA output, \(f_{EOM}\) represents the
frequency modulation by the electro-optic modulator (\(EOM\)), and \(t_{d}\) is the time it takes for
the laser to pass through the delayed line and depends on the length and
the refractive index of the fiber.

In these equations, three different behaviors of the DSHI measurement
are modelled with \(S_{1}\), \(S_{2}\) and \(S_{3}\) parameters.
\(S_{1}\) is a function demonstrating the classical Lorentzian line
the shape of the laser and is independent of the delay line, \(S_{2}\) is
used to model the effect of the coherence behavior in the two lines and
\(t_{d}\) plays a crucial role in it, while the \(S_{3}\) parameter models
the impulse function of the DSHI, which is due to the difference between the two
lines frequency and is only valid when the measurement frequency is very
close to the EOM modulation frequency. Since this region is
insignificant for linewidth determination, we can neglect the \(S_{3}\)
parameter and focus on the \(S_{1}\)and \(S_{2}\) parameters. To
demonstrate the effects of each variable on the resulting spectrum, we
employed computer simulation, whose results are illustrated in
\textbf{Fig S.1}. As can be seen in \textbf{Fig S1(a-b)}, changing
the EOM frequency and optical power doesn't affect the line shape of the
interference patterns and only creates horizontal and vertical shifts in
the frequency and intensity domains, respectively. \(f_{EOM}\)determines
the central frequency of the interference in DSHI method while \(P_{0}\)
governs the power level of the fringes witnessed by the photodetector
without changing its line shape. Furthermore, from \textbf{Fig 3(c)}
we can observe that for a constant delay, changing the linewidth of the
laser changes the power level of the peaks and troughs in a way that
smaller linewidths correspond to higher amplitudes in peaks and troughs
without altering their frequency. On the other hand, by increasing the
delay length for a constant linewidth, as shown in \textbf{Fig S1(d)},
the position of the peaks and troughs changes and their count increases
so that the line shape finds a more Lorentzian-like distribution.

In a practical measurement, the PSD data is collected by an ESA and
displayed with dbm unit. \(f_{EOM}\), \(P_{0}\), and \(t_{d}\) are known
and \(\mathrm{\Delta}f\) is constant value determined by the laser.
Thus, the acquired PSD contains all the required data for linewidth
measurement which needs to be extracted. In the method proposed by Zhao
et al. \cite{zhao2022narrow},   they considered two adjacent extreme points and measured their
magnitude difference (\(\mathrm{\Delta}S\)):
\begin{align*}
\mathrm{\Delta}S 
& = 10\log_{10}S_{P} - 10\log_{10}S_{T} = 10\log_{10}\frac{S\left\lbrack f_{EOM} + \frac{(m + 2)c}{2nL} \right\rbrack}{S\left\lbrack f_{EOM} + \frac{(k + 2)c}{2nL} \right\rbrack} \\ 
&=
10\log_{10}\frac{\left\lbrack {\mathrm{\Delta}f}^{2} + \left\lbrack \frac{(k + 2)c}{2nL} \right\rbrack^{2} \right\rbrack}{\left\lbrack {\mathrm{\Delta}f}^{2} + \left\lbrack \frac{(m + 2)c}{2nL} \right\rbrack^{2} \right\rbrack}\frac{\left\lbrack 1 - exp\left( - 2\pi\frac{nL}{c}\mathrm{\Delta}f \right)\cos\left\lbrack (m + 2)\pi \right\rbrack \right\rbrack}{\left\lbrack 1 - exp\left( - 2\pi\frac{nL}{c}\mathrm{\Delta}f \right)\cos\left\lbrack (k + 2)\pi \right\rbrack \right\rbrack}
\end{align*}
where \(S_{P}\) and \(S_{T}\) are the intensity of the peak and trough
, respectively and \(k\) and \(m\ \)are their numbers respectively. If we
consider two adjunct peaks and throughs, \(|m - k| = 1\) and, with all the
parameters known, we can have an equation between \(\mathrm{\Delta}S\)
and \(\mathrm{\Delta}f\). By having the \(\mathrm{\Delta}S\) value from
the measurement, the amount of \(\mathrm{\Delta}f\) can be estimated.
\begin{figure}[htbp]
\centering
\fbox{\includegraphics[width=0.8\linewidth]{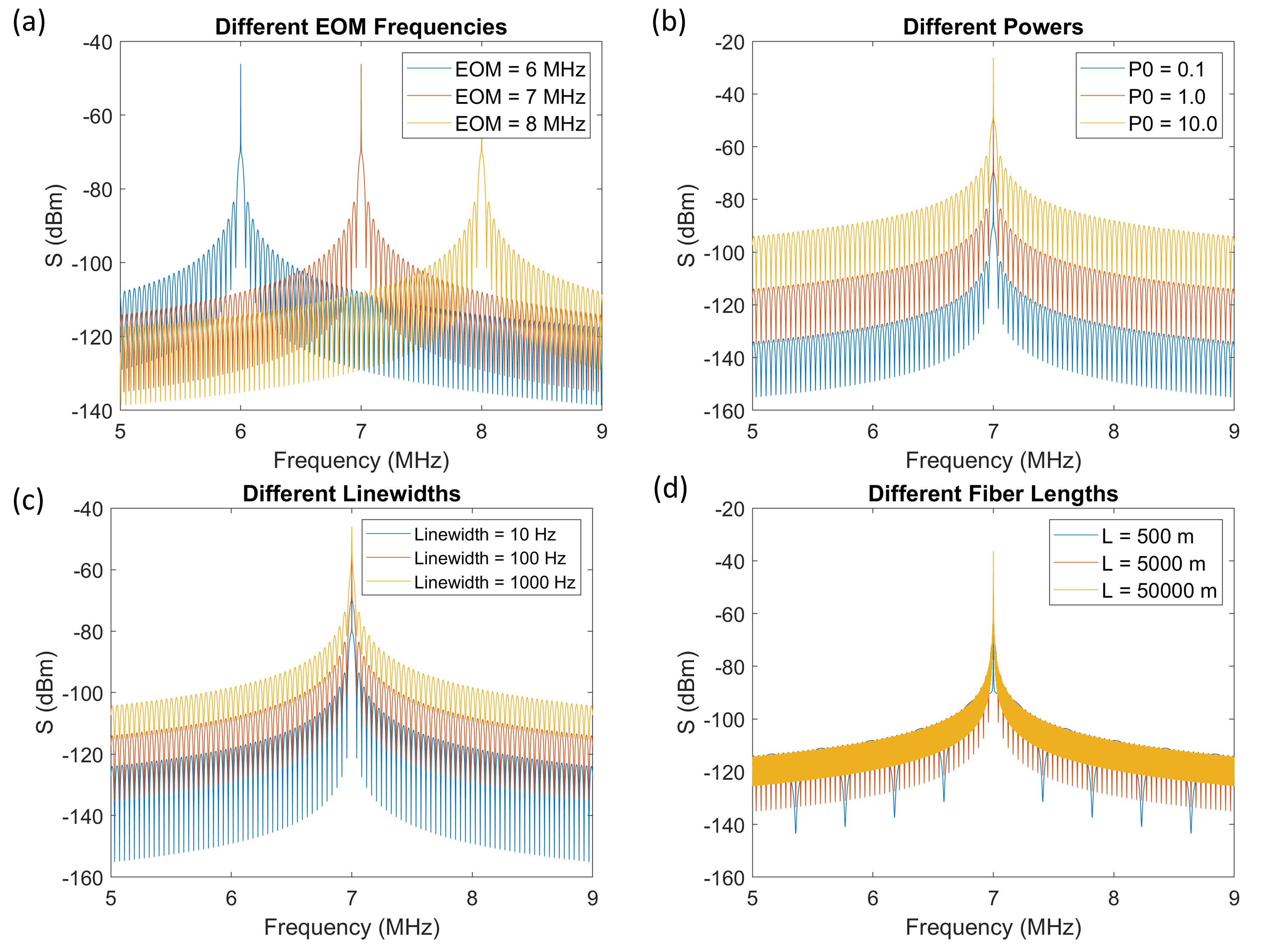}}
\caption{Simulated power spectra for \textbf{(a)} different EOM
frequencies (\(\mathrm{\Delta}f = 100\ Hz\), \(P_{0} = 1\), and
\(L = 5\ km\)), \textbf{(b)} different incident optical power
(\(\mathrm{\Delta}f = 100\ Hz\), \(f_{EOM} = 7\ MHz\), and \(L = 5\ km\)
), \textbf{(c)} different linewidths (\(f_{EOM} = 7\ MHz\),
\(P_{0} = 1\), and \(L = 5\ km\)), and \textbf{(d)} different delay
times or fiber lengths (\(\mathrm{\Delta}f = 100\ Hz\), \(P_{0} = 1\),
\(f_{EOM} = 7\ MHz\)).}

\end{figure}
The mentioned method has exhibited a high capability in estimating narrow linewidths; however, it manifests shortcomings in non-ideal scenarios. In our study, a PID system was employed to introduce servo bumps into the beat note spectrum of the DSHI measurement. Notably, the Huang et al. model does not account for these servo bumps, rendering the reliable measurement of peak contrast ratios challenging. Even if we attempt to model intensity fluctuations caused by the servo bumps, the dependence of servo bandwidths on PID and locking settings introduces further complexity and hinders reliable measurement.

The actual beat note obtained from the DSHI setup is depicted in Fig S2(a), accompanied by its simulated counterpart in Fig S2(c). Both plots reveal peaks at identical frequencies, indicating a successful simulation-experiment alignment in this parameter. However, observed differences in intensity or power spectral densities arise due to their location within the servo bandwidth of the ECDL, as illustrated in Fig S2(b). As previously mentioned, while it is conceivable to incorporate servo bumps into the model, the dependency of their bandwidth on PID settings precludes the attainment of reproducible results.

\begin{figure}[htbp]
\centering
\fbox{\includegraphics[width=0.6\linewidth]{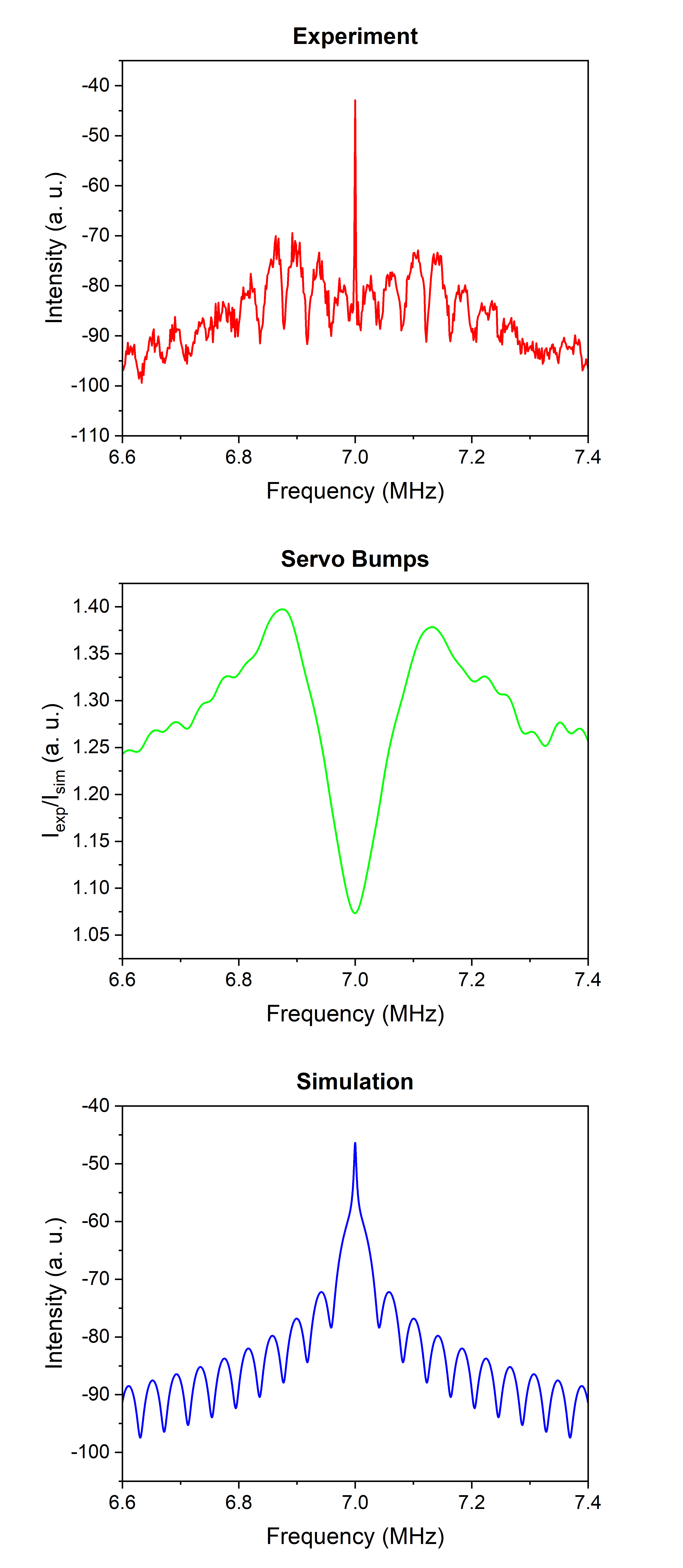}}
\caption{(a) The measured beat note from the TDFA, (b) servo bumps acquired by dividing the experimental data by simulation data, and (c) simulated beat note using the model proposed by Huang et al}

\end{figure}

\begin{figure}[htbp]
\centering
\fbox{\includegraphics[width=0.6\linewidth]{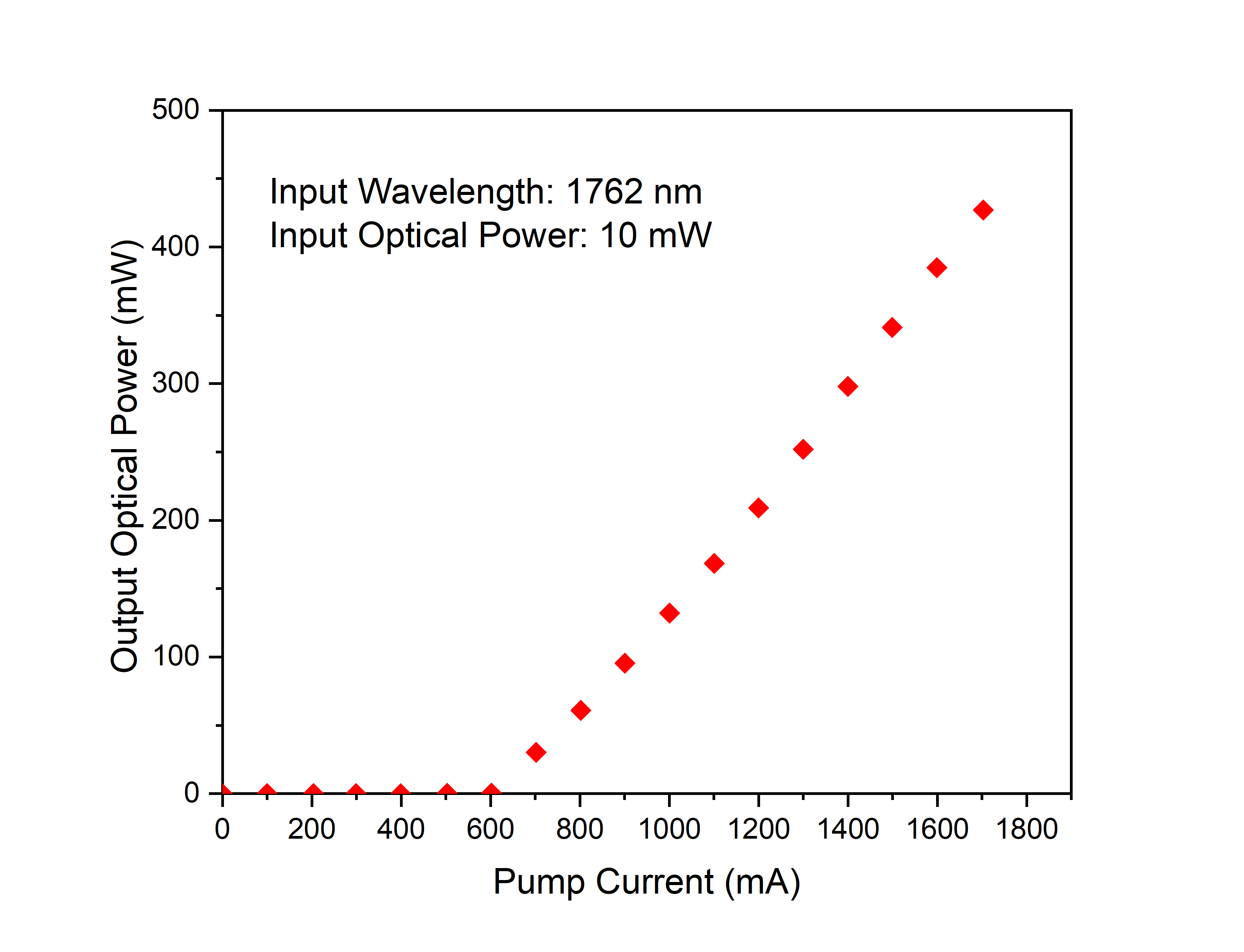}}
\caption{The measured TDFA output amplified power vs its input current. The pump current is responsible for providing the necessary energy to initiate the amplification process by populating the excited states of thulium ions. The amplification gain is not only determined by the pump current but also the input optical power, its polarization and the thermal cooling process of the TDFA. In our measurements, the TDFA unit was cooled down to room temperature using a 3 w fan connected to an aluminum heat sink, and the input polarization was adjusted to match the fast axis of the TDFA  }

\end{figure}
\begin{figure}[htbp]
\centering
\fbox{\includegraphics[width=0.6\linewidth]{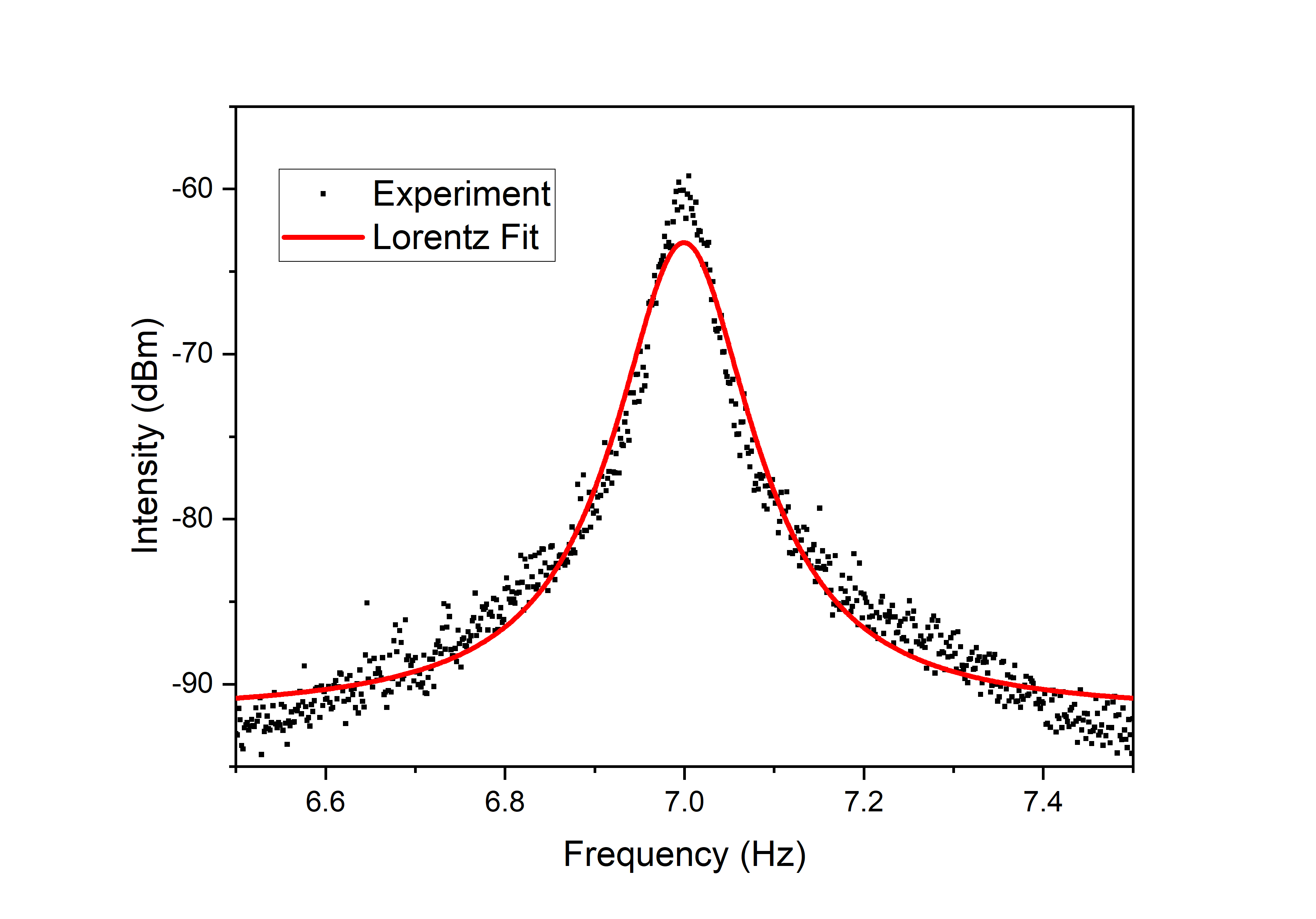}}
\caption{The Delayed Self Heterodyne Interferometry (DSHI) beat note spectrum of the TDFA when the input External Cavity Diode Laser (ECDL) is not locked. In this situation, the linewidth of the laser is broadened to around 100 kHz and the 5-km delay line is sufficient to measure the Lorentzian linewidth.  }

\end{figure}

\begin{figure}[t]
\centering
\fbox{
\includegraphics[width=0.45\textwidth]{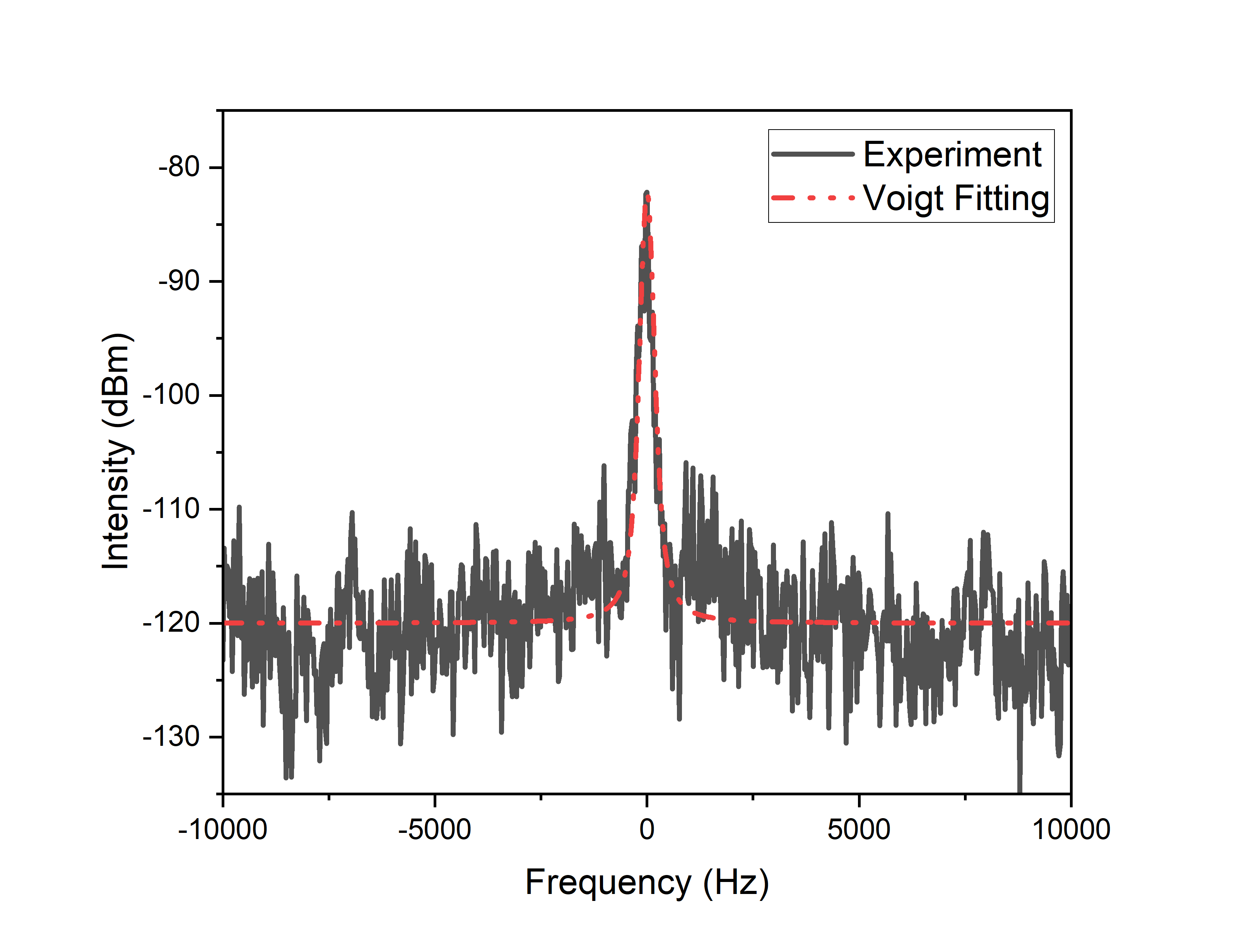}
\includegraphics[width=0.45\textwidth]{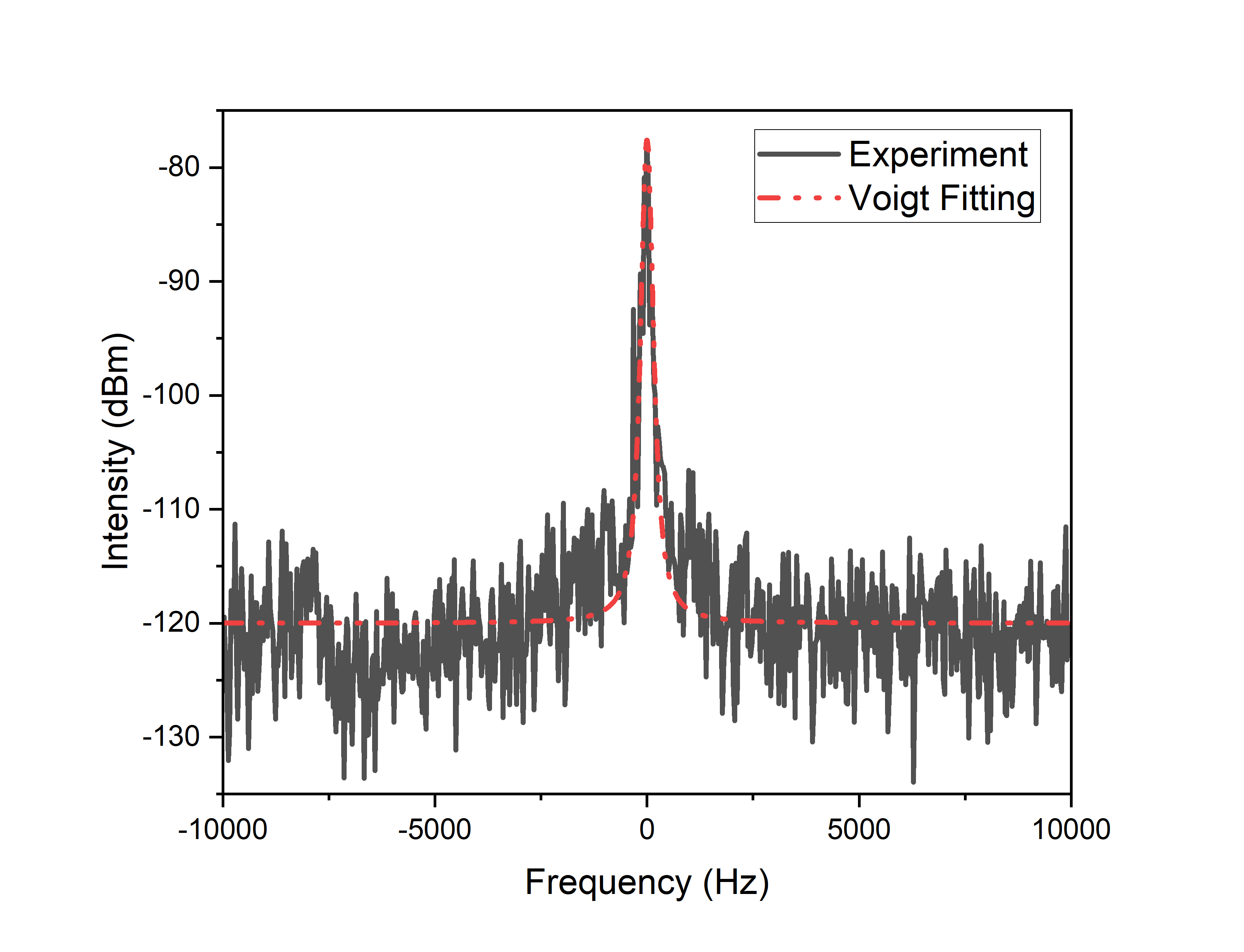}

}
\caption{ Investigating the consistency of TDFA linewidth by tuning the master laser. The DSHI beat note and the fitted Voigt profile for frequency of the master laser at (left) 170124.9 THz and (right) 170123.4 THz, with their frequencies respectively 1.5 and 3.0 GHz away from the laser frequency used in high-resolution atomic spectroscopy. In both cases, the Lorentzian linewidth is estimated to be around 160 Hz, which shows excellent consistency with the master laser and demonstrates the of TDFA in maintaining the linewidth of the input laser while amplifying its power in NIR range }
\end{figure}


\end{document}